\documentclass[twocolumn, showpacs, prl, amsmath, amssymb, amsthm, superscriptaddress, english]{revtex4-1}

\usepackage{graphicx}
\usepackage{color}

\input{Qcircuit}

\begin{document}
\title{Multi-qubit gates protected by adiabaticity and dynamical decoupling
applicable to donor qubits in silicon }
\author{Wayne M. \surname{Witzel} \email{wwitzel@sandia.gov}}
\affiliation{Center for Computing Research, Sandia National Laboratories, Albuquerque, New Mexico 87185 USA}
\author{In\`es \surname{Monta\~no}}
\affiliation{Sandia National Laboratories, New Mexico 87185 USA}
\author{Richard P. \surname{Muller}}
\affiliation{Center for Computing Research, Sandia National Laboratories, Albuquerque, New Mexico 87185 USA}
\author{Malcolm S. \surname{Carroll}}
\affiliation{Sandia National Laboratories, Albuquerque, New Mexico 87185 USA}

\begin{abstract}
We present a strategy for producing multi-qubit gates that promise high fidelity with
minimal tuning requirements.  Our strategy combines gap protection from the adiabatic 
theorem with dynamical decoupling in a complementary manner.  Energy-level transition
errors are protected by adiabaticity and remaining phase errors are mitigated via dynamical 
decoupling.  This is a powerful way to divide and conquer the various error
channels. In order to accomplish this without violating a No-Go theorem
regarding black-box dynamically corrected gates [Phys. Rev. A 80, 032314 (2009)], we require a robust
operating point (sweet spot) in control space where the qubits interact with little sensitivity 
to noise.  There are also energy gap requirements for effective adiabaticity.
We apply our strategy to an architecture in Si with P donors where we assume we can shuttle electrons between different donors.  
Electron spins act as mobile ancillary qubits and P nuclear spins act as long-lived data 
qubits. This system can have a very robust operating point where the electron spin is bound 
to a donor in the quadratic Stark shift regime.  High fidelity single qubit gates may 
be performed using well-established global magnetic resonance pulse sequences.  
Single electron spin preparation and measurement has also been demonstrated.
Putting this all together, we present a robust universal gate set for quantum computation.
\end{abstract}

\maketitle

One of the main challenges in realizing a quantum information processor is the ability 
to implement high-fidelity entangling operations.  It can be relatively easy to control 
well isolated qubits.  Nuclear magnetic resonance (NMR) and electron spin resonance (ESR) are well developed for manipulating nuclear and electron spins with high fidelity~\cite{vandersypen_nmr_2005, morton_high_2005}.  
Turning interactions between qubits on and off in a controllable manner for a coherent quantum operation remains very challenging.  
The process of coupling different qubits is often accompanied by an enhanced
sensitivity to the environment.  When qubits are not isolated, they are vulnerable to noise.

The adiabatic theorem~\cite{kato_adiabatic_1950} provides remarkably robust operations in the sense that transitions 
between non-degenerate eigenstates are suppressed.  If the Hamiltonian of a quantum system 
is varied slowly enough, instantaneous eigenstates will be tracked.  
Exploiting this phenomonon, dramatic improvements in single-qubit NMR operations have been observed~\cite{sigillito_fast_2014} by combining the BIR-4~\cite{garwood_symmetric_1991} pulse sequence with WURST-20~\cite{kupce_adiabatic_1995} adiabatic pulse shaping.
In our proposal, we perform an adiabatic process 
involving two qubits and non-degenerate eigenstates (diabatic energy level crossings are allowed, however).  
Transition errors are suppressed by the adiabatic theorem, but phase errors must be 
mitigated using a different mechanism.

The Hahn echo \cite{hahn_spin_1950} is simple and effective for canceling phase errors induced by 
low-frequency noise and uncertainty.  An unknown but systematic $\hat{Z}$ rotation on a qubit is reversed by flipping
it with an $\hat{X}$ gate.
The Hahn echo, and a variety of more elaborate sequences or 
strategies~\cite{carr_effects_1954, meiboom_modified_1958, khodjasteh_fault-tolerant_2005,uhrig_keeping_2007, viola_robust_2003}
are very effective at prolonging coherence and storing quantum information.
These are known as dynamical decoupling (DD) schemes because they decouple the qubit system from its environment.
There exist analogous strategies called dynamically corrected gates (DCGs) to cancel errors during nontrivial quantum gate operations \cite{khodjasteh_dynamically_2009, khodjasteh_dynamical_2009, khodjasteh_arbitrarily_2010, green_high-order_2012, paz-silva_general_2014}.  
However, a No-Go theorem forbids black-box DCGs~\cite{khodjasteh_dynamical_2009}, presenting a challenge relative to DD sequences.
DCGs must assume there are relationships between the effects of noise induced under different control settings.  In a two-qubit DCG, for example, 
you would need to vary the inter-qubit interaction but maintain consistent or 
correlated environmental interactions in order to cancel their effects.  This presents a 
problem when interactions are varied by moving the qubits (such as localized electrons in a solid state material) and the environment varies 
at this length scale.

\begin{figure}
\includegraphics[width=\linewidth]{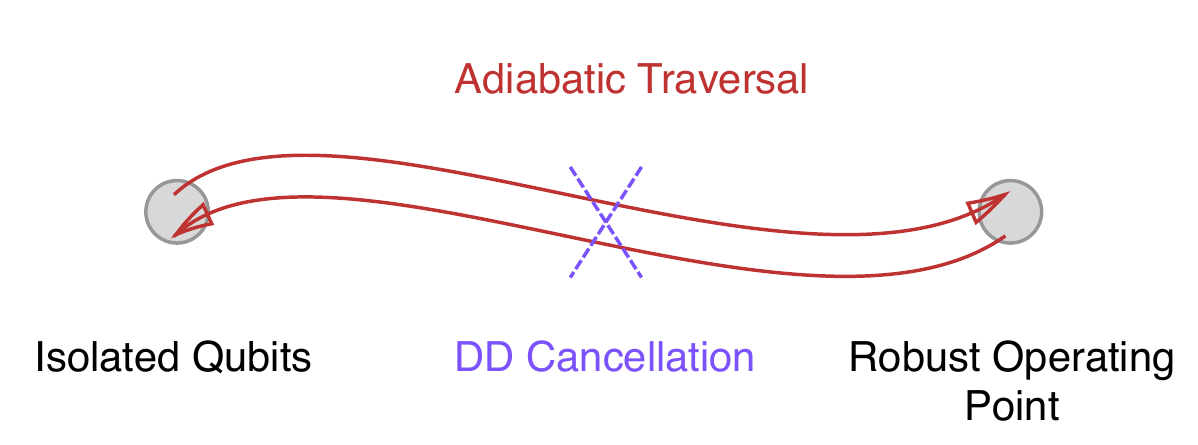} 
\caption{\label{Fig:ControlSpaceSchematic}Control space schematic showing an
adiabatic path between isolated qubits and a ROP where the qubits interact.  We
will cancel phase accumulated during this traversal using dynamical decoupling. }
\end{figure}

We demonstrate a way to circumvent this No-Go theorem when there exists a robust operating
point (ROP), a "sweet spot" in control space where the qubits interact stably with respect to
noise (already exploited in various semiconductor qubit settings~\cite{laird_coherent_2010, koh_high-fidelity_2013, pla_high-fidelity_2013, muhonen_storing_2014}).
We do not attempt to correct for errors induced during this ROP time,  but we do
correct for errors induced in transit (adiabatically) to and from this control space point.  
This is illustrated schematically in Fig.~\ref{Fig:ControlSpaceSchematic}.  We will show 
how this can be accomplished in a generic model in three nested components, prove that 
it provides a universal gate set when combined with single qubit operations,
and then discuss the suitability of silicon donor qubits for implementing this scheme.

The first of our nested components is the adiabatic cycle of moving isolated qubits to a ROP, where they interact, and then back.  
In an ideal limit, adiabatic operations are, by
definition, diagonal with respect to instantaneous eigenstate bases.
Up to an irrelevant global phase such an operation for two qubits is generically
\begin{equation}
\label{Eq:DiagReps}
\begin{array}{ccc}
\left[
\begin{array}{cccc}
e^{i \alpha} & 0 & 0 & 0 \\
0 & e^{i \beta} & 0 & 0 \\
0 & 0 & e^{i \gamma} & 0 \\
0 & 0 & 0 & e^{-i (\alpha + \beta + \gamma)}
\end{array}
\right]
& \equiv &
\begin{array}{c}
\Qcircuit @C=1em @R=.7em {
& \gate{Z_{a}} & \ctrl{1} & \qw\\
& \gate{Z_{b}} & \gate{Z_{c}} & \qw 
}
\end{array}
\end{array}
\end{equation}
in the eigenstate basis using a matrix representation (left) or a circuit-model
representation (right). 
We assume that there is good energy gap protection throughout, including when the qubits are isolated, 
so that low frequency noise will only induce phase errors.  That is, we assume we are not $T_1$ limited.
Consider the path illustrated in 
Fig.~\ref{Fig:ControlSpaceSchematic} consisting of three stages: 1) traversing 
from isolated qubits to the ROP; 2) waiting at the ROP; 3) traversing back to isolated qubits.  Each is an operation 
as in Eq.~(\ref{Eq:DiagReps}) and these operations commute.  Generically, 
we may write the circuit model representation as
\begin{eqnarray}
\nonumber
\begin{array}{c}
\Qcircuit @C=1em @R=.7em {
& \gate{Z_{a_1}} & \ctrl{1} & \gate{Z_{a_2}} & \ctrl{1} & \gate{Z_{a_3}} & \ctrl{1} & \qw\\
& \gate{Z_{b_1}} & \gate{Z_{c_1}} &  \gate{Z_{b_2}} & \gate{Z_{c_2}} &  \gate{Z_{b_3}} & \gate{Z_{c_3}} & \qw 
}
\end{array} \\
\label{Eq:AdiabaticCycle}
=
\begin{array}{c}
\Qcircuit @C=1em @R=.7em {
& \gate{Z_{a}} & \ctrl{1} & \gate{Z_{d(\tau)}} & \ctrl{1} & \qw\\
& \gate{Z_{b}} & \gate{Z_{c}} &  \gate{Z_{e(\tau)}} & \gate{Z_{f(\tau)}} & \qw 
}
\end{array}
\end{eqnarray}
where $a = a_1+a_3$, $b = b_1 + b_3$, $c = c_1 + c_3$, $d(\tau) = a_2$, $e(\tau) = b_2$, and 
$f(\tau) = c_2$.  
We parameterize this operation with $\tau$, the amount of time spent at the ROP.
Somewhat arbitrarily, we will refer 
to the top rail of Eq.~(\ref{Eq:AdiabaticCycle}) as the ancilla and the bottom rail as the data.

The $d(\tau)$, $e(\tau)$, and
$f(\tau)$ phases are incurred at the ROP and therefore presumed to have little sensitivity to noise.
In contrast, the $a$, $b$, and $c$ phases are sensitive to noise and uncertainty during transit.
The second and third components of our procedure are designed to cancel these dependences.  
In the second component, we cancel out the $a$ dependence by applying DD to the 
ancilla qubit in what we will call an ``ancilla-refocused double-cycle''.
This component has three stages.  First, we perform an adiabatic cycle where we set $\tau$ 
such that $f(\tau) = \pi$.  Next, we apply a refocusing $X$ gate on the ancilla qubit.  
Finally, we perform another adiabatic cycle that is the same as the first except we take $\tau=0$.
It is only important that the difference in the two adiabatic cycles amounts to an
extra controlled-Z operation (it is unnecessary to literally spending zero time at the ROP for one of the cycles).
To the extent these operations are not ideally realized, there will be noise that 
is not canceled, but it is instructive to assume idealism in the initial analysis.  
The net operation of this component is then
\begin{equation}
\label{Eq:ancillaDDcPhase}
\begin{array}{c}
\Qcircuit @C=0.5em @R=.3em {
& \gate{Z_{a}} & \ctrl{1} & \gate{Z_{d}} & \control \qw & \gate{X} & \gate{Z_{a}} & \ctrl{1} & \qw \\
& \gate{Z_{b}} & \gate{Z_{c}} &
\gate{Z_{e}} & \ctrl{-1} & \qw  & \gate{Z_{b}} & \gate{Z_{c}} & \qw
}
\end{array}
=
\begin{array}{c}
\Qcircuit @C=0.5em @R=.3em {
& \control \qw & \gate{Z_{d}} & \gate{X} & \qw \\
& \ctrl{-1} & \gate{Z_{g}} & \qw & \qw
}
\end{array}
\end{equation}
where $g = 2 b + c + e$.  Notice that, in addition to canceling the $a$ dependence, we have also made the $c$ rotation on the data qubit deterministic rather than dependent upon the ancilla state.
As far as the ancilla qubit is concerned, it has performed a CPhase operation with the data qubit and the uncertainty has been canceled.

The final, top-level component applies DD in a three stage process as before,
but it will involve two different ancilla qubits.  In the first stage, we do 
an ancilla-refocused double-cycle with one ancilla.  Next, we perform a refocusing 
$X$ gate on the data qubit.  Finally, we do another ancilla-refocused double-cycle 
but with a different ancilla.  The circuit-model representation is
\begin{equation}
\label{Eq:DataRefocus}
\begin{array}{c}
\Qcircuit @C=0.5em @R=.3em {
& \control \qw & \gate{Z_{d}} & \gate{X} & \qw & \qw \\
& &  & \gate{Z_{d}} & \control \qw & \gate{X} & \qw \\
& \ctrl{-2} & \gate{Z_{g}} & \gate{X} & \ctrl{-1} & \gate{Z_{g}} & \qw
}
\end{array}
=
\begin{array}{c}
\Qcircuit @C=0.5em @R=.3em {
& \control \qw & \gate{Z_{d}} & \gate{X} & \qw & \qw \\
& \qw & \gate{Z_{d}} & \control \qw & \gate{X} & \qw \\
& \ctrl{-2} & \gate{X} & \ctrl{-1} & \qw & \qw
}
\end{array}
\Rightarrow
\begin{array}{c}
\Qcircuit @C=0.5em @R=.3em {
& \control \qw & \qw & \qw \\
& \qw  & \control \qw & \qw \\
& \ctrl{-2} & \ctrl{-1} & \qw
}
\end{array}
\end{equation}
where the bottom rail is the data qubit.  We permuted commuting operators on the left side 
of Eq.~(\ref{Eq:DataRefocus}) for compactness.  The right side of the equation shows how 
the $g$ dependence (which is a function of the uncertain $b$ and $c$ parameters) is canceled.
On the right of the arrow is the equivalent operation up to single qubit operations that are 
perfectly known in the ideal limit with respect to being adiabatic, repeatable, insensitive 
to noise at the ROP, and with ideal single-qubit operations.  
We can understand non-ideal operations as evolving states into superpositions of ideal and errant parts,
limiting the overall gate fidelity.

The robust multi-qubit gate that we have produced, shown at the right of Eq.~(\ref{Eq:DataRefocus}), 
is not standard.  We produce two CPhase gates between one data qubit two different ancilla.  
To prove that this is a sufficient for
universality, we show how to produce a single CPhase gate between two
data qubits, mediated by ancilla.  First, it is straightforward to produce a
CPhase between an ancilla and data by discarding one of the ancilla:
\begin{equation}
\label{Eq:ReduceToCPhase}
\begin{array}{c}
\Qcircuit @C=0.5em @R=.3em {
& \control \qw & \qw & \qw & \qw \\
\lstick{\ket{0}} & \qw & \control \qw & \qw & \meter \\
& \ctrl{-2} & \ctrl{-1} & \qw & \qw
}
\end{array}
=
\begin{array}{c}
\Qcircuit @C=0.5em @R=.3em {
& \control \qw & \qw \\
& \ctrl{-1} & \qw
}
\end{array}.
\end{equation}
This is wasteful, but suffices for a proof.
With this data-ancilla CPhase, we can produce a data-data CPhase via
\begin{equation}
\label{Eq:Universality}
\begin{array}{c}
\Qcircuit @C=0.4em @R=.3em {
& \lstick{\ket{0}} & \gate{H} & \ctrl{1} & \gate{H} & \ctrl{2} & \gate{H} & \ctrl{1} & \gate{H} & \meter \\
& \qw & \qw & \control \qw & \qw & \qw & \qw & \control \qw & \qw & \qw \\
& \qw & \qw & \qw & \qw & \control \qw & \qw & \qw & \qw & \qw
}
\end{array}
=
~~
\begin{array}{c}
\Qcircuit @C=0.4em @R=.3em {
& \lstick{\ket{0}} & \targ & \ctrl{2} & \targ & \meter \\
& \qw & \ctrl{-1} & \qw & \ctrl{-1} & \qw \\
& \qw & \qw & \control \qw & \qw & \qw
}
\end{array}
=
\begin{array}{c}
\Qcircuit @C=0.4em @R=.7em {
& \ctrl{1} & \qw \\
& \control \qw & \qw
}
\end{array}.
\end{equation}
Also note that data measurements may be performed indirectly from ancilla measurements using
\begin{equation}
\label{Eq:DataMeasure}
\begin{array}{c}
\Qcircuit @C=0.5em @R=.3em {
\lstick{\ket{0}} & \gate{H} & \control \qw & \gate{H} & \meter \\
& \qw & \ctrl{-1} & \qw & \qw
}
\end{array}.
\end{equation}
Preparation can be implemented via measurement.
Thus, along with full single-qubit control of data qubits, Hadamard gates, and
measurement on ancilla qubits, our multi-qubit operation forms a universal gate set.

We now transition from an abstract to a concrete proposal applied to donor
qubits in silicon.  We envision a similar layout as the well-known 
Kane architecture \cite{kane_silicon-based_1998} in which we have array of P donors in
Si and donor electrons are controlled with electrostatic pads from above. 
Rather than mediating interactions through the exchange coupling of electrons,
however, we propose to shuttle individual electrons between donors as proposed in
Refs.~\onlinecite{skinner_hydrogenic_2003} and
\onlinecite{morton_silicon-based_2009}, possibly by shuttling the electron along an
oxide interface \cite{calderon_quantum_2006, calderon_external_2007}.
The innovation in our proposal is the use of adiabaticity and DD to
cancel uncertainty and low-frequency noise incurred during the shuttling
process.  We treat the electron spins as ancilla qubits and donor nuclear spins
as data qubits and apply our robust multi-qubit gate proposal directly to this
system.  Single electron spin preparation may be performed via spin-selective tunneling into a single-electron transistor~\cite{elzerman_single-shot_2004, morello_single-shot_2010, simmons_tunable_2011}.  Single qubit operations can be performed using global ESR and NMR~\cite{vandersypen_nmr_2005, morton_high_2005}.  
We can implement selective data qubit operations by addressing only donors that are
occupied with properly initialized electrons. Universality does not require
ancilla gate operations to be selective beyond the shuttling done with local
electrostatic controls; the ancilla only need to be able to mediate data qubit
interactions selectively (and in parallel).  The
two-qubit interaction is simply the hyperfine (HF)  coupling between an electron
and the donor it is occupying.  

In order to establish the suitability of our multi-qubit gate strategy to this
Si:P system, we must address the following questions.  How isolated are the
qubits when the interaction is supposed to be off?  How robust is the ROP? 
How adiabatic can we make the shuttling process?


\begin{figure}
\includegraphics[width=\linewidth]{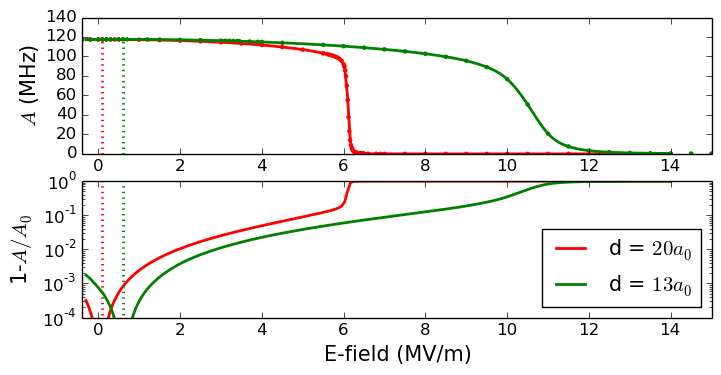} 
\caption{\label{Fig:HfVsEfield} Top: Hyperfine coupling ($A$) computed in
NEMO-3D as a function of E-field for a P donor in Si near an interface at two
different depths ($a_0 \approx 0.54~{\rm nm}$ is the lattice constant).  
Bottom:
Fractional HF difference from its maximum.
Dotted line indicates the ROP (sweet spot).}
\end{figure}

When electrons and nuclei are sufficiently far apart, the dipolar interaction is
the dominant coupling.  When wavefunctions of electrons overlap, the exchange
interaction dominates.  When the electron wavefunction has considerable
amplitude on a phosphorus donor, the contact HF interaction dominates.  We'll assume the
dipolar interaction dominates in the regime in which we regard qubits to be
isolated.  The dipolar Hamiltonian between a pair of spins is
\begin{equation}
H_{\rm D} = \frac{\mu_0 \gamma_1 \gamma_2 \hbar^2}{4 \pi r^3} \left[\vec{\bf{I}}_1 \cdot \vec{\bf{I}}_2 -  \frac{3 (\vec{\bf{I}}_1 \cdot \vec{r})  (\vec{\bf{I}}_2 \cdot \vec{r})}{r^2} \right]
\end{equation}
where $\gamma_1$ and $\gamma_2$ are respective gyromagnetic ratios, $\vec{\bf
I}_1$ and $\vec{\bf I}_2$ are respective spin operators, and $\vec{r}$ is the
vector between the spin positions (the sign is unimportant).  Using appropriate 
gyromagnetic ratios for an electron ($\gamma_S = g \mu_B$ with $g \approx 2$) and 
P nucleus ($\gamma_P = 10.8 \times 10^7 / T s$), this interaction strength is at 
most $105~\rm{MHz} / (r/\rm{nm})^3$ between electrons, $64~\rm{kHz} / (r/\rm{nm})^3$ 
between an electron and nuclear spin, and $40~\rm{Hz} / (r/\rm{nm})^3$ between nuclei.
Nuclear data qubits are well isolated from each other (tens of $\mu$Hz interaction strength 
at $100$~nm).  Electron ancilla qubits can also be regarded as well isolated if their lifetimes
between measurements are short compared to 10~ms ($100~\rm{Hz}$ interaction scale at 
$100$~nm electron separation and $10$~nm electro-nuclear separation).

The contact HF between an electron (ancilla) and nucleus (data)
dominates the inter-qubit interaction.  The Hamiltonian for one electron and
nucleus is
\begin{equation}
\label{Eq:hfHam}
\hat{\cal H} = B \left(\gamma_S {\bf S}^z - \gamma_P {\bf I}^z\right) + A(\vec{E}) {\bf S} \cdot {\bf I}
\end{equation}
where ${\bf S}$ and ${\bf I}$ are respective electron and nuclear spin operators, $B$ is 
the magnetic field (B-field) applied along $z$, and $A$ is the HF interaction
that is electric field (E-field) dependent.
Figure~\ref{Fig:HfVsEfield} shows $A$ versus E-field that we computed in NEMO-3D as in Ref.~\onlinecite{rahman_high_2007}.  The bottom figure shows a 
log-scale view of the ROP at the maximum of $A$.  NEMO-3D predicts
relatively large minimum orbital energy gaps of about
$0.9$~meV and $0.3$~meV for the $13a_0$ and $20a_0$ cases respectively, not
expected to limit the adiabatic transfer rate~\cite{calderon_quantum_2006, calderon_external_2007}.
We neglect anisotropic
HF~\cite{hale_calculation_1971} with the P nucleus (unknown at non-zero
E-field), HF with $^{29}$Si~\cite{witzel_electron_2010}, spin-orbit
interactions~\cite{rahman_gate-induced_2009}, and
g-factor variations~\cite{rahman_high_2007} as sub-dominant effects to be studied in future work.

\begin{figure}
\includegraphics[width=\linewidth]{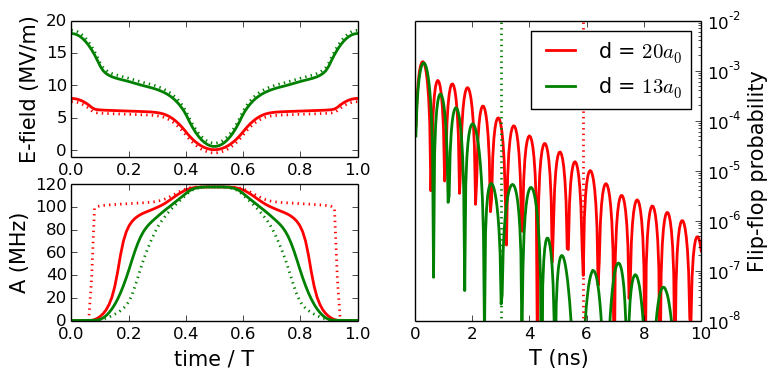} 
\caption{\label{Fig:ShuttleAdiabaticity}Left: Heuristic electron shuttling
schedule with respect to E-field (top) and HF interaction $A$ (bottom) for neutralizing then ionizing a donor near an interface corresponding to Fig.~\ref{Fig:HfVsEfield}.  Dotted curves show hypothetical E-field shifts of $\Delta E = \pm 0.3~$MV/m as considered in Fig.~\ref{Fig:DriftSensitivity}.  Right: Flip-flop failure probability as a function of shuttling time in a $100 $~mT B-field, showing the timescale required to be adiabatic.  The vertical dotted lines indicate shuttle times used in Fig.~\ref{Fig:DriftSensitivity}.}
\end{figure}

To test adiabaticity, we used a heuristic control schedule, shown on the left of 
Fig.~\ref{Fig:ShuttleAdiabaticity}, that limits the first and second order time
derivatives of the E-field and $A$ (motivated by 
findings of Ref.~\onlinecite{lidar_adiabatic_2009} that adiabaticity improves by 
setting time derivatives of the initial and final Hamiltonian to zero). 
While the schedule could be optimized further, our simple heuristic already performs very well as shown on the right of 
Fig.~\ref{Fig:ShuttleAdiabaticity} based upon simulations using 
QuTiP~\cite{johansson_qutip_2013, johansson_qutip:_2012}.  In a $100~$mT field
(or more), the probability of non-adiabatic electro-nuclear flip-flops is very
low for shuttle times of a few nanoseconds.

\begin{figure}
\includegraphics[width=\linewidth]{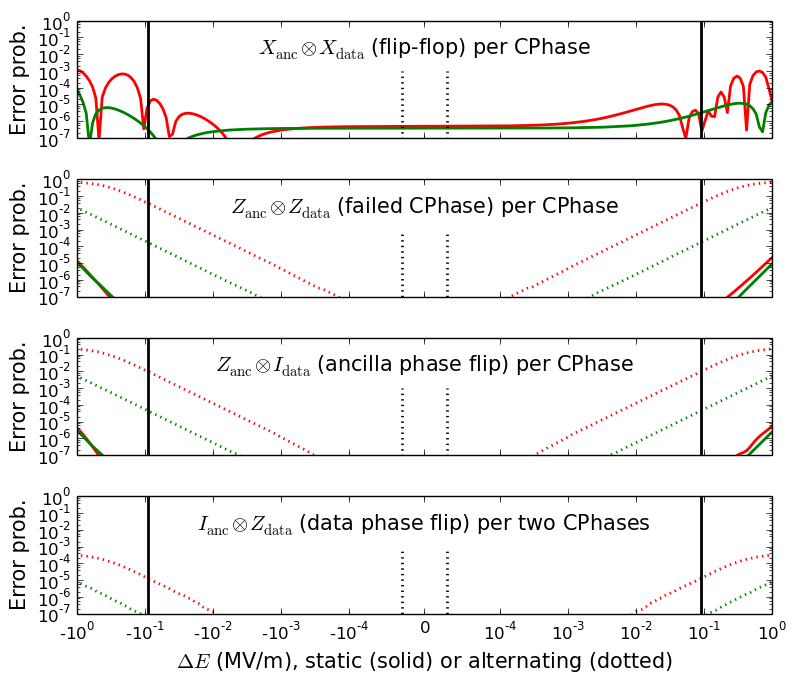} 
\caption{\label{Fig:DriftSensitivity}(Color online) Sensitivity to static (solid) or alternating (dotted) shifts in the E-field, $\Delta E$, corresponding
 to schedules indicated in Fig.~\ref{Fig:ShuttleAdiabaticity} for a donor $20a_0$ (red) or $13a_0$ (green) from an interface.  
Each panel is for a different  
noise channel.  Static shifts probe the low-frequency noise limit.  Alternating shifts probe the worst-case sensitivy to E-field fluctuations with a $\pm \Delta E / 2$ shift before/after the relevant DD refocusing pulse.  Vertical lines indicate magnitude estimates of slow (solid), $0.1$ day scale, and fast (dotted), millisecond scale (to be conservative), fluctuations.  These are based upon charge drift observations reported in Ref.~\onlinecite{zimmerman_charge_2014} 
($0.15 \times 22~\rm{mV} = 3.3~\rm{mV}$), E-field versus voltage for a Si qubit device in Ref.~\onlinecite{laucht_electrically_2015} ($0.026~\rm{MV/m}$ per mV), and an assumption that the noise behaves in a diffusive manner like a random walk).}
 \end{figure}

To illustrate the great benefit from the DD that our scheme employs, Fig.~\ref{Fig:DriftSensitivity}
compares sensitivity to static versus dynamic E-field shifts for the various
error channels.  
The error probability for each channel in
Fig.~\ref{Fig:DriftSensitivity} (and right of
Fig.~\ref{Fig:ShuttleAdiabaticity}) is the Born rule probability 
for the worst-case initial state (orthoganol to the error channel).
Static shifts are the limit of low-frequency noise and well 
tolerated because of DD.  
Flip-flop error probabilities (top) increase with abrupt transitions in the shuttling schedule
that result from static E-field shifts.
The remaining error channels are sensitive to static shifts away from the ROP.
Alternating shifts
are the worst-case higher frequency noise, flipping the sign of the 
shift at the time of the refocusing pulse.
This causes differences of the time integration of $A$, increasing error rates for the bottom 
three error channels of Fig.~\ref{Fig:DriftSensitivity}.
The contrast between the solid and dotted curves illustrates the tremendous benefit of DD.  
We calculate very high entangling gate operations, well below error correction thresholds,
from estimates of fast and slow voltage fluctuations/drift based upon actual device observations 
[Refs.~\onlinecite{zimmerman_charge_2014} and~\onlinecite{laucht_electrically_2015}].


In conclusion, we present a procedure for making robust, universal
multi-qubit operations even if qubit-environment interactions change non-trivially 
when inter-qubit interactions turn on and off.  
Our composite sequence is extremely efficient relative to other
DD~\cite{khodjasteh_fault-tolerant_2005} and
DCG~\cite{khodjasteh_arbitrarily_2010, kestner_noise-resistant_2013} strategies,
making it less vulnerable to control noise,
because we exploit the adiabatic theorem to eliminate most error channels.
In order to circumvent a No-Go theorem that prohibits black-box DCGs 
\cite{khodjasteh_dynamical_2009}, we require a sweet spot (ROP) for pairwise
interactions that may be turned on and off adiabatically.
The requirements are well met for a system of P donors in Si, 
using electron and nuclear spins as two species of qubits.
Our calculations estimate remarkable insensitivity to expected low
frequency E-field noise.

We acknowledge numerous discussions with intellectual contributions to this work from our diverse, multi-disciplinary team of quantum device and architecture experts at Sandia National Laboratories including Nathan Bishop, Robin Blume-Kohout, John Gamble, Anand Ganti, Matthew Grace, N. Tobias Jacobson, Andrew Landahl, Erik Nielsen, and Kevin Young.
We also acknowledge R. Rahman and G. Klimeck for assistance and support with the NEMO-3D simulations.

Sandia National Laboratories is a multi-program laboratory operated by  Sandia Corporation, a wholly owned subsidiary of Lockheed Martin Corporation, for the U.S. Department of Energy's National Nuclear Security Administration under contract DE-AC04-94AL85000. 

\bibliography{hfCPhase}

\end{document}